\documentclass[aps,preprint,showpacs,groupedaddress,floatfix]{revtex4}

\usepackage{graphicx}
\usepackage{dcolumn}
\usepackage{bm}

\begin{document}

\title{Stellar electron-capture rates calculated with the
finite-temperature relativistic random-phase approximation}
\author{Y. F. Niu$^1$}
\author{N. Paar$^{2}$}
\author{D. Vretenar$^2$}
\author{J. Meng$^{3,1,4}$}
\email{mengj@pku.edu.cn}
 \affiliation{$^1$State Key Laboratory of
Nuclear Physics and Technology, School of Physics, Peking
University, Beijing 100871, China}
 \affiliation{$^2$Physics
Department, Faculty of Science, University of Zagreb, Croatia}
 \affiliation{$^3$School of Physics and Nuclear Energy Engineering,
Beihang University, Beijing 100191, China}
 \affiliation{$^4$Department of Physics, University of Stellenbosch, Stellenbosch 7602, South Africa}

\date{\today}
\begin{abstract}
We introduce a self-consistent microscopic theoretical framework for
modelling the process of electron capture on nuclei in stellar
environment, based on relativistic energy density functionals. The
finite-temperature relativistic mean-field model is used to
calculate the single-nucleon basis and the occupation factors in a
target nucleus, and $J^{\pi} = 0^{\pm}$, $1^{\pm}$, $2^{\pm}$
charge-exchange transitions are described by the self-consistent
finite-temperature relativistic random-phase approximation. Cross
sections and rates are calculated for electron capture on
$^{54,56}$Fe and $^{76,78}$Ge in stellar environment, and results
compared with predictions of similar and complementary model
calculations.

\end{abstract}
 \maketitle
\date{today}

\section{Introduction}

Weak interaction processes play a crucial role in the late evolution
stages of massive stars by determining the core entropy and
electron-to-baryon ratio $Y_e$, two important quantities associated
with the dynamics of core-collapse supernovae~\cite{Bethe1979}. At
the end of its life, a massive star exhausts the nuclear fuel and,
therefore, the core can only be stabilized by the electron
degeneracy pressure as long as its mass does not exceed the
corresponding Chandrasekhar mass $M_{\rm Ch}$, proportional to
$Y_e^2$. When this mass limit is exceeded, the core cannot attain a
stable configuration and it collapses. During the pre-collapse
phase, electron capture reduces the number of electrons available
for pressure support, whereas beta-decay acts in the opposite
direction. At the same time, the neutrinos produced by electron
capture freely escape from the star for values of the matter density
$\lesssim 10^{11}$ g cm$^{-3}$, removing energy and entropy from the
core~\cite{Bethe1990,Langanke2003,Janka2007}. For initial values of
$Y_e$ around $0.5$, $\beta^-$ decay processes can be effectively
hindered by electron degeneracy, but get to be competitive when
nuclei become more neutron-rich.

For central stellar densities less than a few $10^{10}$ g/cm$^3$ and
temperatures between $300$ keV and $800$ keV, electron capture
mainly occurs on nuclei in the mass region $A \sim 60$. Under such
conditions electron-capture rates are sensitive to the detailed
Gamow-Teller (GT) strength distribution, because the electron
chemical potential is of the same order of magnitude as the nuclear
$Q$-value (defined as the difference between neutron and proton
chemical potentials). For even higher densities and temperature,
nuclei with mass numbers $A>65$ become quite abundant. The electron
chemical potential is noticeably larger than the $Q$-value, thus
electron-capture rates are principally determined by the total GT
strength and its centroid energy. At core densities $\rho > 10^{11}$
g/cm$^3$, the electron chemical potential reaches values larger than
about $20$ MeV, and forbidden transitions can no longer be
neglected~\cite{Langanke2003,Janka2007}.

Because of its relevance in modelling supernovae evolution, the
process of electron capture has been studied employing various
approaches, often based on available data. The first standard
tabulation of nuclear weak-interaction rates for astrophysical
applications was that of Fuller, Fowler and Newman
(FFN)~\cite{Fuller1980,Fuller1982,Fuller1982AJ,Fuller1985}. It was
based on the independent particle model, but used experimental
information whenever available. The tables included rates for
electron capture, positron capture, $\beta$-decay, and positron
emission for relevant nuclei in the mass range $21 \leq A \leq 60$.
The shell model Monte Carlo method (SMMC) was used to determine for
the first time in a fully microscopic way the GT contributions to
presupernova electron-capture rates for fp-shell nuclei, taking into
account thermal effects. The electroweak interaction matrix elements
were calculated in the zero-momentum transfer limit, with the GT
operators as the main ingredient. The GT strength distributions were
obtained from the response function in the canonical ensemble,
solved in the $0\hbar\omega$ fp-shell space~\cite{Dean1998}. The
diagonalization of the correponding Hamiltonian matrix in the
complete pf-shell model space reproduces the experimental GT$^+$
distributions~\cite{Langanke1998PLB,Caurier1999,Langanke2000}. An
updated tabulation of weak interaction rates for more than $100$
nuclei in the mass range $45\leq A \leq 65$, with the same
temperature and density grid as the one reported by FFN, was carried
out based on the large-scale shell-model diagonalization (LSSM)
approach~\cite{Langanke2001data}.

An alternative approach to the calculation of weak-interaction rates
is based on the random-phase approximation (RPA). This framework is
generally more suitable for the inclusion of forbidden transitions,
and for global calculations involving a large number of nuclei
included in nuclear networks. To overcome the limitations of the
shell model, in a study of nuclei beyond the fp-shell a hybrid model
was introduced. In this approach the SMMC is used to obtain the
finite-temperature occupation numbers in the parent nucleus, and the
allowed and forbidden transitions for the electron-capture process
are calculated in the random-phase approximation using mean-field
wave functions with the SMMC occupation numbers~\cite{Langanke2001}.
More recently the hybrid model plus the RPA, with a global
parametrization of single-particle occupation numbers, has been
employed in estimates of electron-capture rates of a large number of
nuclei involved in stellar core collapse~\cite{Juodagalvis.10}.

Electron-capture rates were also calculated for sd-shell and
fpg-shell nuclei using the proton-neutron quasiparticle RPA (QRPA)
approach, based on the Nilsson model and separable GT
forces~\cite{Nabi1999,Nabi2004}. However, the use of  experimental
masses for calculation of $Q$-values limits the application of this
model to nuclei with known masses. More recently a thermal QRPA
approach (TQRPA) has been introduced, based on the Woods-Saxon
potential and separable multipole and spin-multipole particle-hole
interactions, with temperature taken into account using the
thermofield dynamics (TFD) formalism~\cite{Dzhioev2010}. A fully
self-consistent microscopic framework for evaluation of nuclear
weak-interaction rates at finite temperature has recently been
introduced, based on Skyrme density functionals. The single-nucleon
basis and the corresponding thermal occupation factors of the
initial nuclear state are determined in the finite-temperature
Skyrme Hartree-Fock model, and charge-exchange transitions to
excited states are computed using the finite-temperature
RPA~\cite{Paar2009}.

An important class of nuclear structure models belongs to the
framework of relativistic energy density functionals (EDF). In
particular, a number of very successful relativistic mean-field
(RMF) models have been very successfully employed in analyses of a
variety of nuclear structure phenomena, not only in nuclei along the
valley of $\beta$-stability, but also in exotic nuclei with extreme
isospin values and close to the particle drip
lines~\cite{Ring1996PPNP,Meng2006PPNP,Vretenar2005PRE}. Based on
this framework, the relativistic (Q)RPA has been developed and
applied in studies of collective excitations in nuclei, including
giant resonances, spin-isospin resonances, and exotic modes of
excitation in unstable
nuclei~\cite{Ma2001,Ring2001,Niksic2002,Paar2003,Paar2004,Liang2008,PVKC.07}.
By employing a small set of universal parameters adjusted to data,
both ground-state properties and collective excitations over the
whole chart of nuclides, from relatively light systems to superheavy
nuclei, can be accurately described. For studies of astrophysical
processes, temperature effects have recently been included in the
self-consistent relativistic RPA. The low-energy monopole and dipole
response of nuclei at finite temperatures were
investigated~\cite{Niu2009}. An extension of the finite-temperature
relativistic RPA (FTRRPA) to include charge-exchange transitions,
will certainly provide a very useful theoretical tool for studies of
the electron-capture process in presupernova collapse.

In this work we introduce the theoretical framework, based on the
charge-exchange FTRRPA, for the calculation of electron-capture
cross sections and stellar electron-capture rates on selected
medium-mass nuclei. The single nucleon basis and the thermal
occupation factors of the initial nuclear state are determined in a
finite-temperature RMF model, and charge-exchange transitions to the
excited states are computed using the FTRRPA. The same relativistic
energy density functional is consistently used both in the RMF and
RPA equations. The advantage of this approach is that the
calculation is completely determined by a given energy density
functional and, therefore, can be extended over arbitrary mass
regions of the nuclide chart, without additional assumptions or
adjustment of parameters, as for instance single-particle energies,
to transitions within specific shells. In a simple RPA, of course,
correlations are described only on the one-particle -- one-hole
level, and therefore one cannot expect the model to reproduce the
details of the fragmentation of GT strength distributions.

The paper is organized as follows. In Sec. II the framework of the charge-exchange
FTRRPA and the formalism for the electron-capture cross sections and rates are
introduced. The Gamow-Teller strength distributions at finite temperature are
discussed in Sec. III. The calculated electron-capture cross sections and rates
in a stellar environment are presented in Sec. IV and V, respectively.
Sec. VI summarizes the present work and ends with an outlook for
future studies.

\section{Formalism}

Since electron capture on nuclei involves charge-exchange
transitions, for the purpose of the present study we extend the
self-consistent finite-temperature relativistic random-phase
approximation (FTRRPA)~\cite{Niu2009} and implement the model in the
charge-exchange channel. The characteristic properties of the
nuclear initial state, that is, the single-nucleon basis and the
corresponding thermal occupation probabilities, are obtained using
an RMF model at finite temperature. This framework was introduced in
Ref.~\cite{Gambhir2000}, based on the nonlinear effective Lagrangian
with the NL3 parameterization~\cite{Lalazissis1997}. In this work
the RMF at finite temperature is implemented using an effective
Lagrangian with medium-dependent meson-nucleon
couplings~\cite{Typel1999,NVFR.02}. The corresponding FTRRPA
equations are derived using the single-nucleon basis of the RMF
model at finite temperature~\cite{Niu2009}. In a self-consistent
approach the residual interaction terms in the FTRRPA matrix are
obtained from the same Lagrangian. The proton-neutron FTRRPA
equation reads

\begin{equation}
\left(
 \begin{array}{cc}
 A^J_{p np' n'} &  B^J_{p n p'n'} \\
 - B^J_{pnp' n'} & - A^J_{pn p'n'}
 \end{array}\right)\left(
  \begin{array}{c} X^J_{p' n'} \\ Y^J_{ p'n'} \end{array} \right)
   =\omega_\nu
  \left( \begin{array}{c} X^J_{p n} \\ Y^J_{p n}
   \end{array}\right),
\end{equation}
where $ A$ and $ B$ are the matrix elements of the particle-hole
residual interaction,
 \begin{eqnarray}
 \label{RPAmatrix}
A^J_{p np'n'} &=& (\epsilon_P - \epsilon_{
H})\delta_{pp'}\delta_{nn'} + V^J_{pn'np'}(\tilde{u}_p  \tilde{v}_n
\tilde{u}_{p'}  \tilde{v}_{n'} +  \tilde{v}_p \tilde{u}_n
\tilde{v}_{p'}  \tilde{u}_{n'} )(|f_{n'}- f_{p'}|), \\
B^J_{p np'n'} &=& V^J_{p n' np'} (\tilde{u}_p \tilde{v}_n
\tilde{v}_{p'}  \tilde{u}_{n'} + \tilde{v}_p  \tilde{u}_n
\tilde{u}_{p'}  \tilde{v}_{n'} )(|f_{p'}-f_{n'}|).
\label{RPA2}
 \end{eqnarray}
The diagonal matrix elements contain differences of single-particle
energies between particles and holes $\epsilon_P - \epsilon_{ H}$,
and these could be either $\epsilon_p - \epsilon_{\bar n}$ or
$\epsilon_n - \epsilon_{\bar p}$, where $p$, $n$ denote proton and
neutron states, respectively. For a given proton-neutron pair
configuration, the state with larger occupation probability is
defined as a hole state, whereas the other one is a particle state.
In the relativistic RPA, the configuration space includes not only
proton-neutron pairs in the Fermi sea, but also pairs formed from
the fully or partially occupied states in the Fermi sea and the
empty negative-energy states from the Dirac sea. The residual
interaction term $V^J_{pn'np'}$ is coupled to the angular momentum
$J$ of the final state. The spin-isospin-dependent interaction terms
are generated by the exchange of $\pi$ and $\rho$ mesons. Although
the direct one-pion contribution to the nuclear ground state
vanishes at the mean-field level because of parity conservation, the
pion nevertheless must be included in the calculation of
spin-isospin excitations that contribute to the electron-capture
cross section. For the $\rho$-meson density-dependent coupling
strength we choose the same functional form used in the RMF
effective interaction~\cite{NVFR.02}. More details about the
corresponding particle-hole residual interaction are given in
Ref.~\cite{PNVR.04}. The factors $f_{p(n)}$ in the matrix elements
$A$ Eq. (\ref{RPAmatrix}) and $B$ Eq. (\ref{RPA2}), denote the
thermal occupation probabilities for protons and neutrons,
respectively. These factors are given by the corresponding
Fermi-Dirac distribution
 \begin{equation}
  f_{p(n)}= \frac{1}{1+{\rm exp}(\frac{\epsilon_{p(n)} -
   \mu_{p(n)}}{kT})},
 \end{equation}
where $\mu_{p(n)}$ is the chemical potential determined by the conservation
of the number of nucleons $\sum_{p(n)} f_{p(n)} = Z (N)$. The
factors $\tilde{u},\tilde{v}$ are introduced in order to distinguish
the GT$^-$ and GT$^+$ channel, that is
 \begin{eqnarray}
 &&  \tilde{u}_p = 0, \quad \tilde{v}_p = 1, \quad\tilde{u}_n = 1, \quad \tilde{v}_n =0, \text{ when } f_p > f_n \quad(\bar p n), \\
 &&  \tilde{u}_p = 1, \quad \tilde{v}_p = 0, \quad\tilde{u}_n = 0, \quad \tilde{v}_n =1, \text{ when } f_p < f_n \quad(p \bar n).
 \end{eqnarray}
With this definition the FTRRPA matrix is decoupled into two
subspaces for the GT$^-$ and GT$^+$ channels.

The FTRRPA equations are solved by diagonalization, and the
results are the excitation energies $E_\nu$ and the corresponding
forward- and backward-going amplitudes $X^{J\nu}$ and $Y^{J\nu}$,
respectively. The normalization reads
 \begin{equation}
\sum_{pn} [(X^{J\nu}_{pn})^2  -  (Y^{J\nu}_{pn})^2 ] (|f_{p}-f_n|)=
1.
 \end{equation}
The transition strengths for GT$^{\pm}$ operators are calculated using the
relations
 \begin{eqnarray}
 \label{transition}
B_{ J\nu}^{T_-} &=& | \sum_{pn} (X^{J\nu}_{p n} \tilde{u}_p
\tilde{v}_n + Y^{J\nu}_{pn}\tilde{v}_p \tilde{u}_n) \langle p || T^-
|| n \rangle
(|f_{n} -f_p|) |^2,\nonumber\\
B_{ J\nu}^{T_+} &=& | \sum_{pn} (X^{J\nu}_{p n} \tilde{v}_p
\tilde{u}_n + Y^{J\nu}_{pn}\tilde{u}_p \tilde{v}_n) \langle p || T^+
|| n \rangle (|f_{n} -f_p|)|^2,
 \end{eqnarray}
where the spin-isospin operators read:
$T^{\pm} = \sum_{i=1}^A \boldsymbol{\sigma} \tau_{\pm}$.

For the process of electron capture on a nucleus
\begin{equation}
 \label{EC}
 e^- + _Z ^A X_N \rightarrow _{Z-1} ^A X_{N+1}^* + \nu_e,
\end{equation}
the cross section is derived from Fermi's golden rule:
 \begin{equation}
  \frac{d\sigma}{d\Omega}
  =  \frac{1}{(2\pi)^2} V^2  E^2_\nu  \frac{1}{2}
 \sum_{\text{lepton spins}} \frac{1}{2J_i+1}\sum_{M_iM_f}
 | \langle f | \hat{H}_W | i \rangle |^2,
 \end{equation}
 where $V$ is the quantization volume, and $E_\nu$ is the energy of the outgoing electron
neutrino. The weak-interaction Hamiltonian $\hat{H}_W$ of
semileptonic processes is written in the current-current
form~\cite{Walecka1975}
 \begin{equation}
  \hat{H}_W = - \frac{G}{\sqrt{2}} \int d \boldsymbol{x} {\cal J}_\mu(\boldsymbol{x})
  j_\mu(\boldsymbol{x}),
 \end{equation}
 where $j_\mu(\boldsymbol{x})$ and ${\cal J}_\mu(\boldsymbol{x})$ are the
 weak leptonic and hadronic current density operators, respectively.
 The matrix elements of
 leptonic part are evaluated using the standard electroweak model,
 and contain both vector and axial-vector
 components~\cite{Langanke2003}. The hadronic current is obtained by
 using arguments of Lorentz covariance and isospin invariance of
 the strong interaction. The expression for  the electron capture cross sections
 (see Refs.~\cite{Connell1972,Walecka1975} for more details) reads
\begin{eqnarray}
  & &\frac{d \sigma_{\rm ec}}{ d \Omega }
  =  \frac{ G_F^2{\rm cos}^2 \theta_c }{2\pi}
  \frac{F(Z,E_e)}{(2J_i+1)} \nonumber \\
& &\times  \Bigg\{ \sum_{J \geq 1} \mathcal{W}(E_e,E_{\nu})
  \Big\{
  {\left(
1-(\hat{\bm{\nu}} \cdot \hat{\bm{q}})(\bm{\beta} \cdot \hat{\bm{q}})
  \right )}
  \left[ \vert \langle J_f || \hat{\mathcal{T}}_J^{MAG} || J_i
\rangle \vert^2
+ \vert \langle J_f || \hat{\mathcal{T}}_J^{EL} || J_i \rangle
\vert^2 \ \right ] \nonumber \\
& & -  2\hat{\bm{q}} \cdot (\hat{\bm{\nu}} -  \bm{\beta} ) {\rm
Re}\langle J_f || \hat{\mathcal{T}}_J^{MAG} || J_i \rangle \langle
J_f || \hat{\mathcal{T}}_J^{EL}|| J_i \rangle^{*} \Big\}
\nonumber \\
& & +\sum_{J \geq 0 } \mathcal{W}(E_e,E_{\nu}) \Big\{
(1-\hat{\bm{\nu}} \cdot  \bm{\beta} + 2(\hat{\bm{\nu}} \cdot
\hat{\bm {q}})(\bm{\beta} \cdot \hat{\bm{q}})) | \langle J_f ||
\hat{\mathcal{L}}_J || J_i \rangle | ^2 + (1+ \hat{\bm{\nu}} \cdot
\bm{\beta}) | \langle J_f || \hat{\mathcal
{M}}_J || J_i \rangle | ^2 \nonumber \\
& & - 2 \hat{\bm{q}}\cdot (\hat{\bm{\nu}} +  \bm{\beta}) {\rm Re}
\langle J_f || \hat{\mathcal{L}}_J || J_i \rangle \langle J_f ||
\hat{\mathcal{M}}_J|| J_i \rangle^{*} \Big\} \Bigg\} \;,
\label{ec_rate}
\end{eqnarray}
where the momentum transfer $\bm{q}=\bm{\nu}-\bm{k}$ is defined as
the difference between neutrino and electron momenta, $\hat{\bm{q}}$
and $\hat{\bm{\nu}}$ are the corresponding unit vectors, and
$\bm{\beta} = \bm{k}/ E_e$. The energies of the incoming electron
and outgoing neutrino are denoted by $E_e$ and $E_{\nu}$,
respectively. The Fermi function $F(Z,E_e)$ corrects the cross
section for the distortion of the electron wave function by the
Coulomb field of the nucleus~\cite{Kol.03}. The explicit energy
dependence of the cross section is given by the term
\begin{eqnarray}
  \mathcal{W}(E_e,E_{\nu}) = \frac{E_{\nu}^2}{(1+ E_{e}/M_T(1-\hat{\bm{\nu}} \cdot\bm{\beta} ))}\; ,
\end{eqnarray}
where the phase-space factor $(1+ E_{e}/M_T(1-\hat{\bm{\nu}}
\cdot\bm{\beta} ))^{-1}$ accounts for the nuclear recoil, and $M_T$
is the mass of the target nucleus. The nuclear transition matrix
elements between the initial state $|J_i \rangle$ and final state
$|J_f \rangle$, correspond to the charge $\hat{\mathcal{M}}_J$,
longitudinal $\hat{\mathcal{L}}_J$, transverse electric $
\hat{\mathcal{T}}_J^{EL}$, and transverse magnetic
$\hat{\mathcal{T}}_J^{MAG}$ multipole
operators~\cite{Connell1972,Walecka1975}. The initial and final
nuclear states in the hadronic matrix elements are characterized by
angular momentum and parity $J^\pi$. In the present calculation a
number of multipoles contributing to the cross section
Eq.~(\ref{ec_rate}) will be taken into account.

In the electron capture process, the excitation energy of the
daughter nucleus $ _{Z-1} ^A X_{N+1}$ is obtained by the sum of the
RPA energy $E_{\rm RPA}$ given with respect to the ground state of
the parent nucleus and the binding energy difference between
daughter and parent nucleus~\cite{Colo1994}. Thus the energy of the
outgoing neutrino is determined by the conservation relation:
 \begin{equation}
 \label{Enu}
 E_\nu = E_e - E_{\rm RPA} -\Delta_{np},
 \end{equation}
where $E_e$ is the energy of incoming electron, and
$\Delta_{np}=1.294$ MeV is the mass difference between the neutron
and the proton. The axial-vector coupling constant $g_A=-1.0$ is
quenched for all the multipole excitations with respect to its
free-nucleon value $g_A=-1.26$. The reason to consider quenching the
strength in all multipole channels, rather than just for the GT is,
of course, that the axial form factor appears in all four transition
operators in Eq.~(\ref{ec_rate}) that induce transitions between the
initial and final states, irrespective of their multipolarity. The
study based on continuum random phase
approximation~\cite{Kolbe1994,Kolbe2000} showed that there is no
indication of the necessity to apply any quenching to the operators
responsible for the muon capture on nuclei. However, recent
calculations of the muon capture rates based on the
RQRPA~\cite{Marketin2009}, employed on a large set of nuclei, showed
that reducing $g_A$ by $10\%$ for all multipole transitions
reproduces the experimental muon capture rates to better than $10\%$
accuracy.

The electron capture rate is expressed in terms of the cross section Eq.~(\ref{ec_rate})
and the distribution of electrons $f(E_e,\mu_e,T)$ at a given temperature:
\begin{equation}
  \lambda_{\rm ec} = \frac{1}{\pi^2 \hbar^3} \int_{E^0_e}^{\infty}
  p_e E_e \sigma_{ec}(E_e) f(E_e,\mu_e,T)
  d E_e.
 \label{ecrate}
 \end{equation}
$ E_e^0 = max(|Q_{if}|, m_ec^2)$ is the minimum electron energy that
allows for the capture process, that is, the threshold energy for
electrons, where $Q_{if} = -E_{\rm RPA} -\Delta_{np} $. $p_e =
(E_e^2 - m_e^2c^4)^{1/2}$ is the electron momentum. Under stellar
conditions that correspond to the core collapse of a supernova, the
electron distribution is described by the Fermi-Dirac
expression~\cite{Juodagalvis.10}
 \begin{equation}
  f(E_e,\mu_e,T) = \frac{1}{{\rm exp}(\frac{E_e - \mu_e}{kT}) +1}.
  \label{fermidirac}
 \end{equation}
$T$ is the temperature, and the chemical potential $\mu_e$ is
determined from the baryon density $\rho$ by inverting the relation
 \begin{equation}
  \rho Y_e = \frac{1}{\pi^2 N_A} \left( \frac{m_e c}{\hbar}\right)^3
  \int_0^\infty (f_e - f_{e^+} ) p^2 dp,
 \end{equation}
where $Y_e$ is the ratio of the number of electrons to the number of
baryons, $N_A$ is Avogadro's number, and $ f_{e^+}$ denotes
the positron distribution function similar to Eq.~(\ref{fermidirac}),
but with $\mu_{e^+}=-\mu_e$. We assume that the phase space
is not blocked by neutrinos.

\section{Gamow-Teller transition strength at finite temperature}

In this section we present an analysis of Gamow-Teller transition
strength distributions at finite temperature for iron isotopes and
neutron-rich germanium isotopes. The GT$^+$ transition is the
dominant process not only in electron capture on nuclei near the
stability line, but also on neutron-rich nuclei because of the
thermal unblocking effect at finite temperature. Here we employ the
finite-temperature relativistic RPA to calculate the GT$^+$ strength
distribution. At zero temperature, however, pairing correlations
have to be taken into account for open shell nuclei, and thus the
Relativistic Hartree Bogoliubov model and the quasiparticle RPA with
the finite range Gogny pairing force are used in the corresponding
calculations (more details are given in Ref.~\cite{Paar2004}). In
atomic nuclei the phase transition from a superfluid to normal state
occurs at temperatures $T\approx 0.5 -1 $
MeV~\cite{Khan2007,Goodman1981,Goodman1981a,Cooperstein1984} and,
therefore, for the temperature range considered in the present
analysis, the FTRRPA should provide a reasonable description of the
Gamow-Teller transitions and electron-capture rates.

\begin{figure}
\centerline{
\includegraphics[scale=0.35,angle=0]{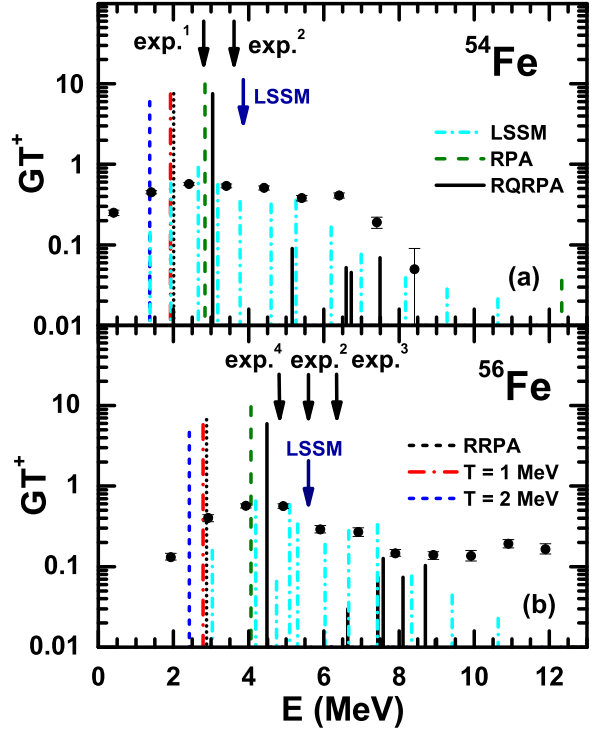}
} \caption{(Color online) The GT$^+$ strength distributions for
$^{54,56}$Fe as functions of the excitation energy with respect to
the ground state of the parent nucleus, calculated with the
proton-neutron RQRPA at zero temperature, and the FTRRPA at $T=0, 1,
2$ MeV, for the DD-ME2 relativistic density functional. For
comparison, the GT$^+$ strength calculated with the non-relativistic
RPA based on the SLy5 Skyrme functional (green dashed lines), and
the centroid energies (blue arrows) and distributions of the LSSM
calculation~\cite{Caurier1999} at $T=0$ MeV are shown. The
experimental centroid energies from
Ref.~\cite{Vetterli1989,Roennqvist1993,El-Kateb1994,Frekers2005} are
indicated by black arrows, and the experimental distributions from
Ref.~\cite{Roennqvist1993} for $^{54}$Fe and
Ref.~\cite{El-Kateb1994} for $^{56}$Fe are shown by solid circles.}
\label{fig1}
\end{figure}

In Fig.~\ref{fig1} we display the GT$^+$ strength distributions for
$^{54,56}$Fe at $T=0, 1, 2$ MeV, as functions of excitation energy
with respect to the ground state of the parent nucleus. At zero
temperature both the RQRPA and RRPA results are shown, whereas the
finite temperature transition spectra are calculated using only the
FTRRPA, that is, pairing is not included in calculations at finite
temperatures. The self-consistent results correspond to the DD-ME2
relativistic density functional~\cite{LNVR.05}. For comparison, the
GT$^+$ strength at zero temperature calculated with the RPA based on
the Skyrme functionals SLy5 parameterization is also shown. The
transition energy is higher and the strength somewhat larger as
compared to the results of the relativistic model. Of course, the
simple (Q)RPA approach cannot reproduce the empirical fragmentation
of the strength, that is, the spreading width. This could only be
accomplished by including additional correlations going beyond the
RPA as, for instance, in the second RPA involving $2p-2h$
configurations~\cite{Drozdz1990}, or in the particle-vibration
coupling model~\cite{Colo1994,Litvinova2007}. The present analysis
is primarily focused on the centroid energy of GT$^+$ transitions,
and model calculations are only carried out on the (Q)RPA level.
Fig.\ref{fig1} also includes the centroid energies and strength
distributions of a large-scale shell model (LSSM)
diagonalization~\cite{Caurier1999}. The experimental centroid
energies~\cite{Vetterli1989,Roennqvist1993,El-Kateb1994,Frekers2005},
defined as the energy-weighted integrated strength over the total
strength, $m_1/m_0$, are indicated by arrows in the figure. The
experimental strength distributions from Ref.~\cite{Roennqvist1993}
for $^{54}$Fe and Ref.~\cite{El-Kateb1994} for $^{56}$Fe are also
shown. The centroid energies and distributions obtained in the LSSM
calculation and the experimental values are displayed with respect
to the ground states of the parent nuclei, for convenience of
comparison with the RPA results.

One might notice that the RQRPA calculation is in fair agreement
with the experimental centroid energies. Compared to the LSSM, the
RQRPA excitation energies are $\approx 1$ MeV lower for both nuclei.
By comparing the RRPA and RQRPA , we notice that pairing
correlations shift the GT$^+$ transition to higher energy by about
$1\sim1.5$ MeV, because additional energy is needed to break a
proton pair. When the temperature is increased to $1$~MeV, the
transition energy is lowered by about $1.1$ MeV for $^{54}$Fe, and
$1.6$ MeV for $^{56}$Fe. This decrease in energy is mainly caused by
the pairing collapse. With a further increase in temperature to $2$
MeV, the GT$^+$ transition energy decreases by about $0.5$ MeV in
both nuclei. This continuous decrease has its origin in the
softening of the repulsive residual interaction because of the
occupation factors that appear in the FTRRPA matrix elements. To
demonstrate this in a quantitative way, we consider the example of
$^{56}$Fe, and analyze the unperturbed energies $E_{\rm unper}$,
that is, the transition energy without residual interaction, and the
energy shift caused by the residual interaction. For $^{56}$Fe the
principal contribution to the GT$^+$ comes from the transition from
the proton orbital $\pi 1f_{7/2}$ to the neutron orbital $\nu
1f_{5/2}$. In the QRPA the unperturbed energy approximately equals
the sum of two quasiparticle energies, and the chemical potential
difference of neutrons and protons, resulting in $E_{\rm
unper}\simeq 3.6$ MeV. The energy shift induced by the repulsive
residual interaction is $0.9$ MeV. If pairing correlations are not
included, that is in RPA, the unperturbed energy corresponds to the
difference between the single-particle energies of the two orbitals,
and this is $1.8$ MeV at zero temperature, and $1.7$ MeV at $T=2$
MeV. Therefore the residual interaction shifts the energy by $1.1$
MeV at zero temperature, and by $0.7$ MeV at $T=2$ MeV. Obviously
the partial occupation factors (the smearing of the Fermi surface),
induced either by pairing correlations or by temperature effects,
will lead to the weakening of the residual interaction. The
temperature effect appears to be more pronounced because the Fermi
surface is more diffuse at $T=2$ MeV. In addition to the excitation
energy, the transition strength could also be reduced by the
smearing of the Fermi surface through the occupation factors in
Eq.~(\ref{transition}). Therefore, the transition strength becomes
weaker with increasing temperature or with the inclusion of pairing
correlations. We have verified that the Ikeda sum
rule~\cite{Ikeda1963} is satisfied at finite temperature.

\begin{figure}
\centerline{
\includegraphics[scale=0.35,angle=0]{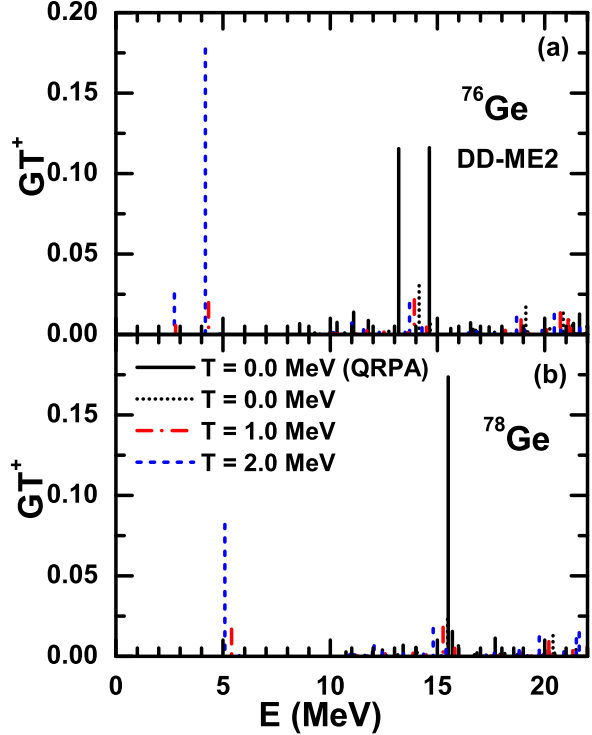}
} \caption{(Color online) The GT$^+$ strength distributions of
$^{76,78}$Ge, calculated with the proton-neutron RQRPA at $T=0$ MeV,
and with the FTRRPA at $T=0, 1, 2$~MeV, using the DD-ME2
relativistic density functional. } \label{fig2}
\end{figure}

In Fig.~\ref{fig2} we plot the GT$^+$ strength distributions of the
neutron-rich nuclei $^{76,78}$Ge at $T=0, 1, 2$ MeV.  At zero
temperature results obtained with both the RQRPA and the FTRRPA are
shown. It is found that almost no transition strength appears at
zero temperature without the inclusion of pairing correlations,
because the GT$^+$ transition channels are Pauli-blocked for these
neutron-rich nuclei. As shown in the figure, the transition channels
can be unblocked by two mechanisms, that is, by pairing correlations
or thermal excitations. Two unblocked single-particle transitions
principally contribute to the total GT$^+$ strength: the $\pi
1g_{9/2} \rightarrow \nu 1g_{7/2}$ particle-particle, and the $\pi
1f_{7/2} \rightarrow \nu 1f_{5/2}$ hole-hole transitions, where
particle (hole) denotes a state above (below) the chemical
potential.

Let us consider $^{76}$Ge as an example, and analyze its evolution
behavior with temperature. With the inclusion of pairing
correlations at $T=0$ MeV, two major peaks are calculated at
$E=15.8$ MeV and $E=16.9$ MeV. The first state mainly corresponds to
the transition $\pi 1f_{7/2} \rightarrow \nu 1f_{5/2}$, whereas the
higher state results from a superposition of the transitions $\pi
1g_{9/2} \rightarrow \nu 1g_{7/2}$ and $\pi 1f_{5/2} \rightarrow \nu
2f_{7/2}$. At $T=1$ MeV the GT$^{+}$ excitations shift to $E=2.8$
MeV and $E=4.3$ MeV, and correspond to the transitions $\pi 1f_{7/2}
\rightarrow \nu 1f_{5/2}$ and $\pi 1g_{9/2} \rightarrow \nu
1g_{7/2}$, respectively, with very weak transition strength. When
the temperature is further increased to $T=2$ MeV, the excitation
energies are only slightly lowered (by $0.1$ MeV), but the
transition strengths are considerably enhanced.

The shift in energy from $T=0$ MeV with pairing correlations, to
$T=1$ MeV is about 13 MeV. This cannot be explained solely by the
removal of the extra energy needed to break a proton pair. To
explain this result, we analyze the unperturbed transition energies.
It is found that the unperturbed energies are much higher when
pairing correlations are included, as compared with the effect of
finite temperature, resulting in considerable difference between the
corresponding GT$^+$ energies. However, it is not only the pairing
gaps that raise the unperturbed energy because, for instance, the
pairing gaps for $\pi 1g_{9/2}$ and $\nu 1g_{7/2}$ are both about
$1.8$ MeV. As these unblocked channels are particle-particle or
hole-hole transitions, the sum of the quasiparticle energies $E_{\rm
qp} = \sqrt{(\epsilon_p -\lambda_p)^2+ \Delta_p^2 } +
\sqrt{(\epsilon_n -\lambda_n)^2+ \Delta_n^2 }$ is much larger than
the difference of the single-particle energies
$\epsilon_n-\epsilon_p$, that corresponds to the unperturbed
energies at finite temperature. This decrease of GT$^+$ excitation
energies is in accordance with the results of
Ref.~\cite{Dzhioev2010}.

The large difference between the RQRPA GT$^+$ strength at $T=0$ MeV
and the FTRRPA strength at $T=1$ MeV is mainly caused by the diffuseness
of the  Fermi surface induced by pairing correlations at zero temperature.
With a further increase of temperature to $T=2$ MeV, the Fermi
surface becomes more diffuse, and this leads to enhancement of the
GT$^+$ strength. A similar trend with temperature increase is found
when the nucleus becomes even more neutron-rich
(cf. the case of $^{78}$Ge in Fig.~\ref{fig2}), but the transition channels are
more difficult to unblock by thermal excitations, and this result in a weaker
transition strength. In the present calculation for $^{78}$Ge only the
particle-particle channel $\pi 1g_{9/2} \rightarrow \nu 1g_{7/2}$ is
unblocked at finite temperature.

To test the sensitivity of the results to the choice of the
effective interaction, we have also carried out the same
calculations for $^{54,56}$Fe and $^{76,78}$Ge using the
relativistic density-dependent effective interaction
PKDD~\cite{Long2004}. The same general behavior is found with both
interactions, but with PKDD the excitation energies are
systematically larger by about $0.5$ MeV for Fe, and by $0.3$ MeV
for the Ge isotopes, whereas the transition strengths are slightly
enhanced compared to the DD-ME2 results.

\section{Electron-capture cross sections}
In this section we calculate electron-capture cross sections for
selected medium-mass target nuclei using RQRPA at zero temperature,
and the FTRRPA at temperatures $T=0, 1,$ and 2 MeV.
\begin{figure*}
\centerline{
\includegraphics[scale=0.37,angle=0]{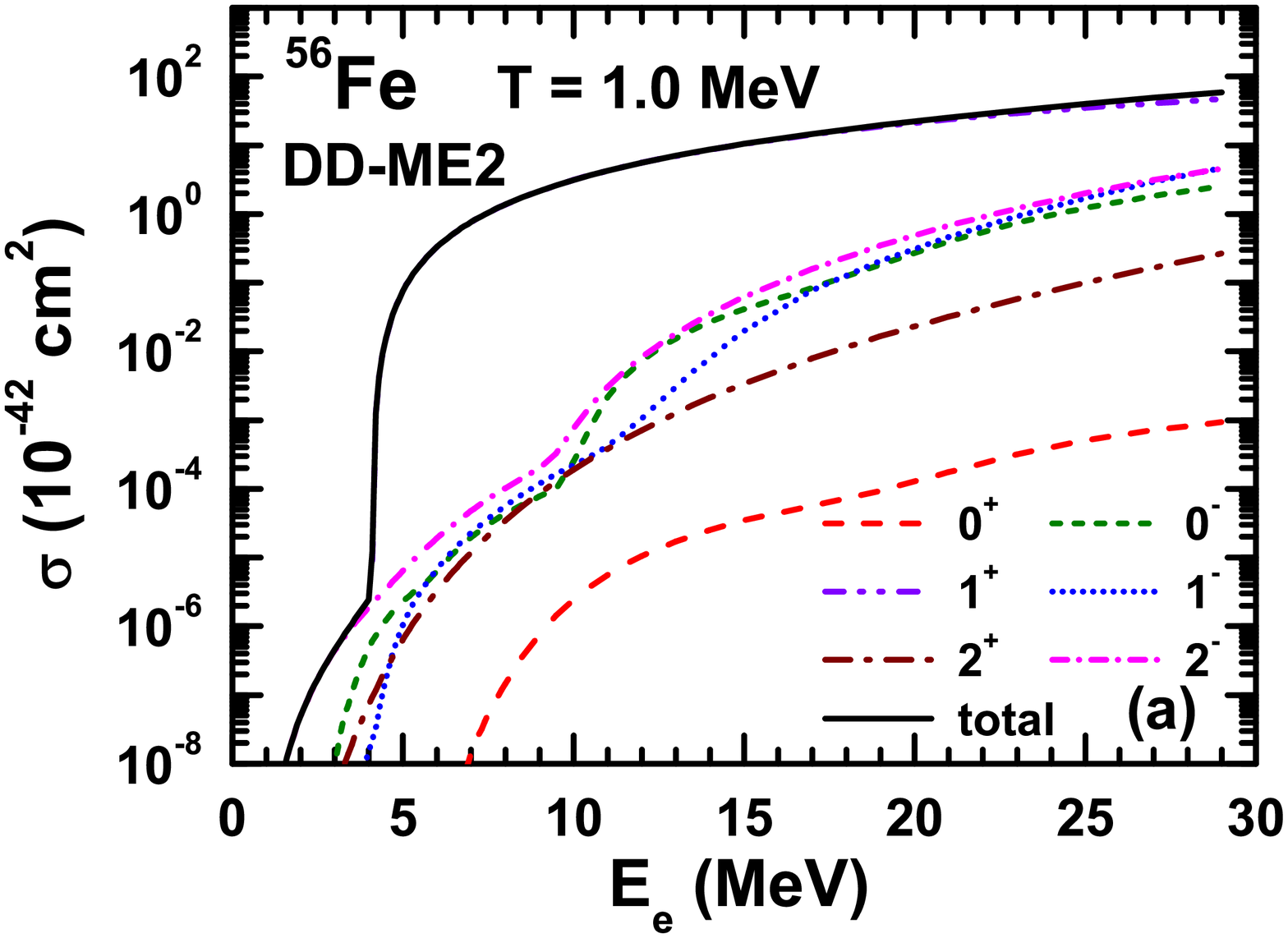}
\includegraphics[scale=0.37,angle=0]{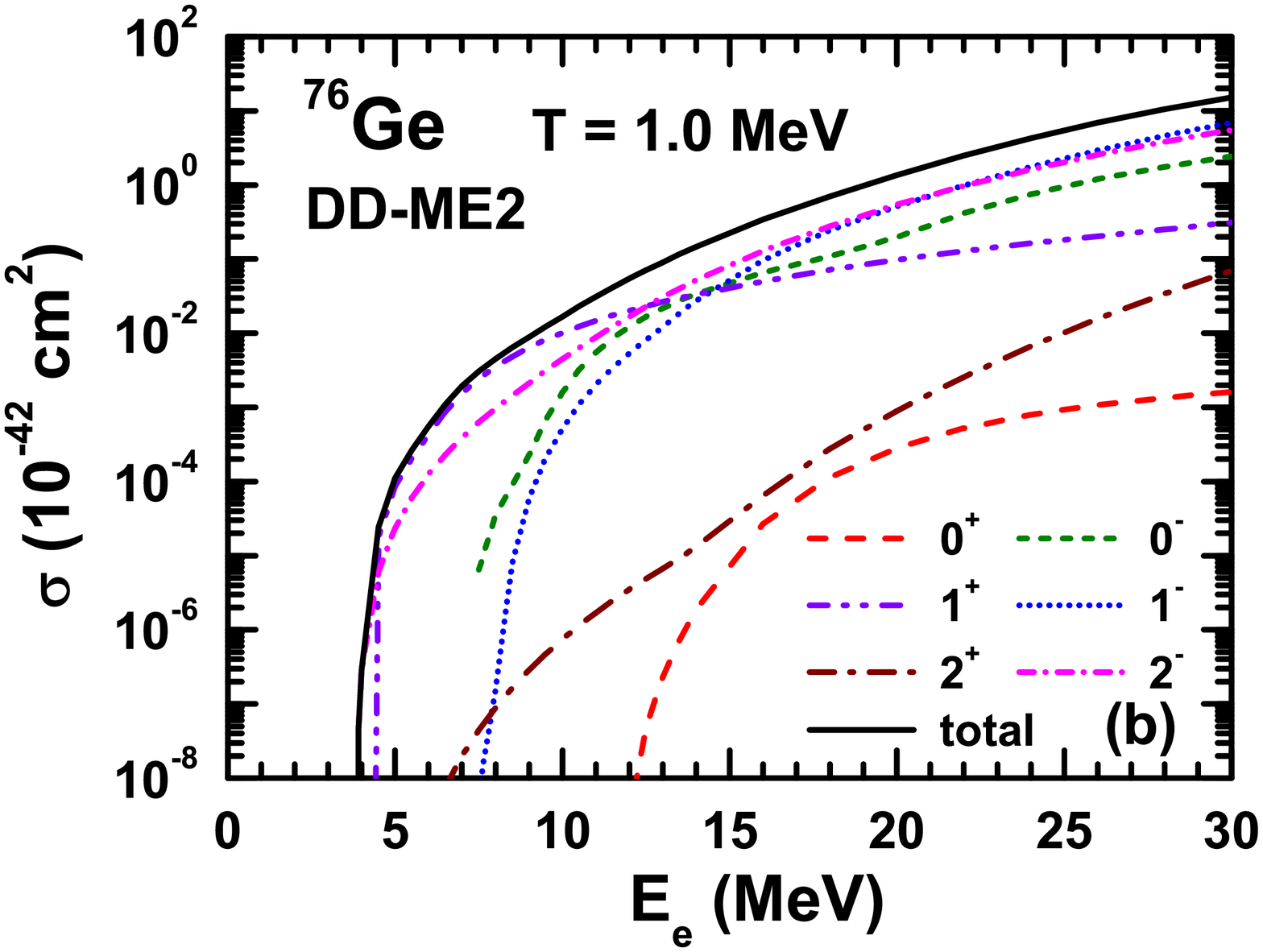}
} \caption{(Color online) Electron-capture cross sections for the
$^{56}$Fe and $^{76}$Ge target nuclei at $T=1$ MeV, calculated with
the FTRRPA using the DD-ME2 effective interaction. In addition to
the total cross section which includes multipole transitions
$J^{\pi} = 0^{\pm}$, $1^{\pm}$, and $2^{\pm}$, contributions from
the individual channels are shown in the plot as functions of the
incident electron energy $E_e$.} \label{fig3}
\end{figure*}
In Fig.~\ref{fig3} the cross sections for electron capture on
$^{56}$Fe and $^{76}$Ge at $T=1$ MeV are plotted as functions of the
incident electron energy $E_e$. The cross sections are calculated
using the expression of Eq.~(\ref{ec_rate}), and the FTRRPA with the
DD-ME2 relativistic density functional~\cite{LNVR.05} is used to
evaluate the transition matrix elements. In addition to the total
cross sections which include multipole transitions $J^{\pi} =
0^{\pm}$, $1^{\pm}$, and $2^{\pm}$, contributions from the
individual channels are shown in the plot, as functions of the
incident electron energy $E_e$. For $^{56}$Fe the total cross
section is completely dominated by the $1^+$ channel (GT$^+$) all
the way up to $E_e=30$ MeV, with contributions from other channels
being orders of magnitude smaller. In the case of the neutron-rich
nucleus $^{76}$Ge, on the other hand, forbidden transitions play a
more prominent role, already starting from $E_e > 12$ MeV. Their
contribution to the total cross section further increases with the
electron energy $E_e$. Obviously in systematic calculations of
electron capture rates on heavier, more neutron-rich nuclei,
contributions from forbidden transitions should also be included in
addition to the GT$^+$ channel.

\begin{figure}
\centerline{
\includegraphics[scale=0.45,angle=0]{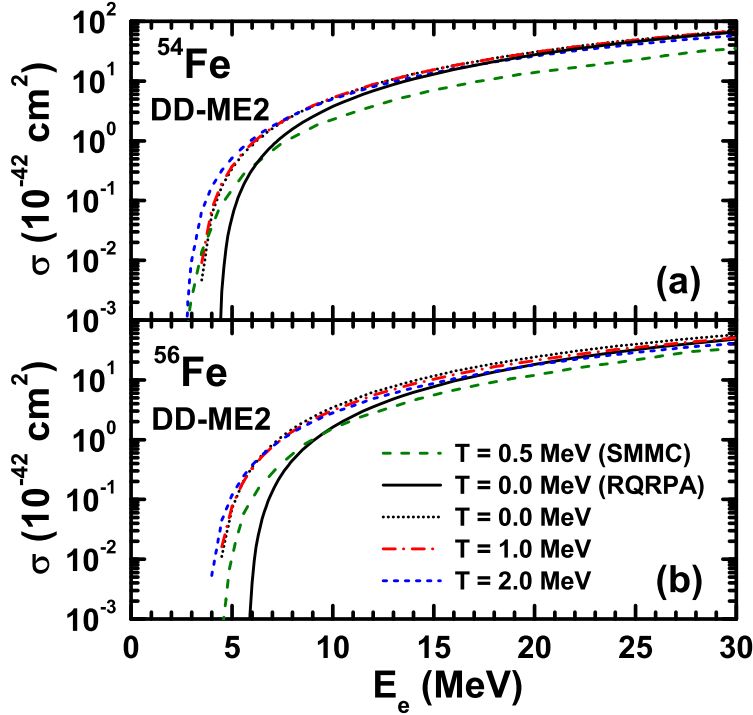}
} \caption{(Color online) Electron-capture cross sections for the
target nuclei $^{54,56}$Fe at $T=0, 1,$ and $2$ MeV, as functions of
the incident electron energy $E_e$. The results obtained with the
proton-neutron RQRPA at $T=0$ MeV, and with the FTRRPA at $T=0, 1,$
and 2 MeV, using the DD-ME2 effective interaction, are shown in
comparison with cross sections calculated from the SMMC GT$^+$
strength distributions~\cite{Dean1998}. } \label{fig4}
\end{figure}

Next we illustrate how the capture cross sections evolve with
temperature. Fig.~\ref{fig4} displays the electron-capture cross
sections for the target nuclei $^{54,56}$Fe at $T=0, 1$, and 2 MeV,
as functions of the incident electron energy $E_e$. Since for
$^{54,56}$Fe forbidden transitions in the range of electron energy
up to 30 MeV give negligible contributions to the total cross
section (cf. Fig.~\ref{fig3}), here only the 1$^+$ transitions are
included in the calculation. Results obtained with the
proton-neutron RQRPA at $T=0$ MeV, and with the FTRRPA at $T=0, 1,$
and 2 MeV, using the DD-ME2 effective interaction, are shown in
comparison with cross sections calculated from the SMMC GT$^+$
strength distributions~\cite{Dean1998}. Note, however, that in the
SMMC calculation only the $0\hbar \omega$ Gamow-Teller transition
strength is considered, rather than the total strength in the $1^+$
channel. We notice that the principal effect of increasing the
temperature in this interval is the lowering of the electron-capture
threshold energy. From $T=0$ MeV (RQRPA) to $T=1$ MeV (FTRRPA) this
decrease is more pronounced than the one from $T=1$ to $2$ MeV, in
accordance with the behavior of GT$^+$ distributions discussed in
the previous section. At low electron energy below 10 MeV one
notices a pronounced difference between the RQRPA and FTRRPA
results, reflecting the treatment of pairing correlations at zero
temperature. Of course, the calculated cross sections become almost
independent of temperature at high electron energies. The results of
the present calculation are in qualitative agreement with those of
the SMMC model~\cite{Dean1998}, calculated at temperature T=0.5 MeV.
Cross sections calculated at very low electron energies are
sensitive to the discrete level structure of the Gamow-Teller
transitions and, therefore, one expects that the SMMC approach will
produce more accurate results. These cross sections, however, are
orders of magnitude smaller than those for $E_e \geq 10$ MeV and,
when folded with the electron flux to calculate capture rates, the
differences between values predicted by various models in the
low-energy interval will not have a pronounced effect on the
electron-capture rates.

\begin{figure}
\centerline{
\includegraphics[scale=0.45,angle=0]{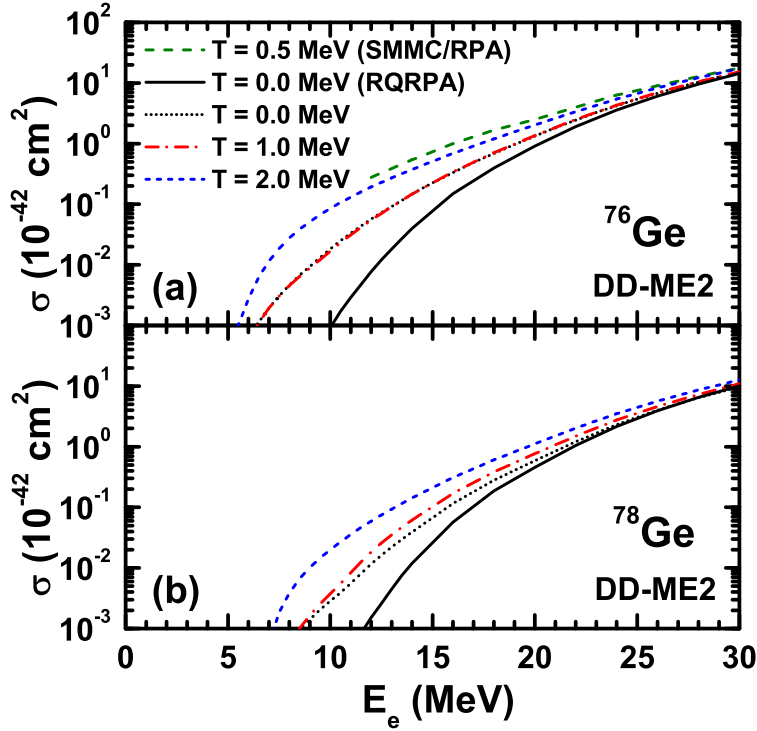}
} \caption{(Color online) Electron-capture cross sections for the
target nuclei $^{76,78}$Ge at $T=0, 1$, and 2 MeV, as functions of
the incident electron energy $E_e$. The results are obtained by
employing the DD-ME2 effective interaction in the proton-neutron
RQRPA at $T=0$ MeV, and in the FTRRPA at $T=0, 1,$ and 2 MeV. For
$^{76}$Ge the results are also compared with the cross section
obtained in the hybrid model (SMMC/RPA) at $T=0.5$ MeV
\cite{Langanke2001}.} \label{fig5}
\end{figure}

In Fig.~\ref{fig5} we also illustrate the temperature dependence of
the electron-capture cross sections for the neutron-rich nuclei
$^{76,78}$Ge. The calculation includes the multipole transitions
$J^\pi=0^{\pm},1^\pm, 2^\pm$. For $^{76}$Ge the results are also
compared with the cross section obtained in the hybrid model
(SMMC/RPA) at $T=0.5$ MeV~\cite{Langanke2001}. One might notice that
the cross sections are reduced by about an order of magnitude when
compared to the Fe isotopes, but overall a similar evolution with
temperature is found. By increasing the temperature the threshold
energy for electron capture is reduced. The cross sections exhibit a
rather strong temperature dependence at electron energies $E_e \leq
12$ MeV. At $E_e = 12$ MeV, by increasing the temperature by $1$
MeV, the cross sections are enhanced about half an order of
magnitude. Since at $E_e \leq 12$ MeV the electron capture
predominantly corresponds to  GT$^+$ transitions (see
Fig.~\ref{fig3}), the enhancement of the cross sections is caused by
the thermal unblocking of the GT$^+$ channel, similar as predicted
by the hybrid SMMC/RPA model~\cite{Langanke2001}. For higher
electron energies, forbidden transitions become more important. The
results of the present analysis are in qualitative agreement with
those of the TQRPA model calculation~\cite{Dzhioev2010}, and the
finite-temperature RPA approach based on Skyrme
functionals~\cite{Paar2009}. It is also found that the hybrid
model~\cite{Langanke2001} predicts slightly larger cross sections at
lower energies, as anticipated due to the strong configuration
mixing in SMMC calculations. In general, by increasing the number of
neutrons in target nucleus, the electron capture occurs with a
higher threshold and smaller cross sections.


\section{Stellar electron-capture rates}

In modelling electron-capture rates in stellar environment one
assumes that the atoms are completely ionized, and the electron gas
is described by the Fermi-Dirac distribution (\ref{fermidirac}). By
folding the FTRRPA cross sections at finite temperature with the
distribution of electrons in Eq.~(\ref{ecrate}), we calculate the
rates for electron capture on Fe and Ge isotopes, under different
conditions associated with the initial phase of the core-collapse
supernova.

\begin{figure}
\includegraphics[scale=0.45,angle=0]{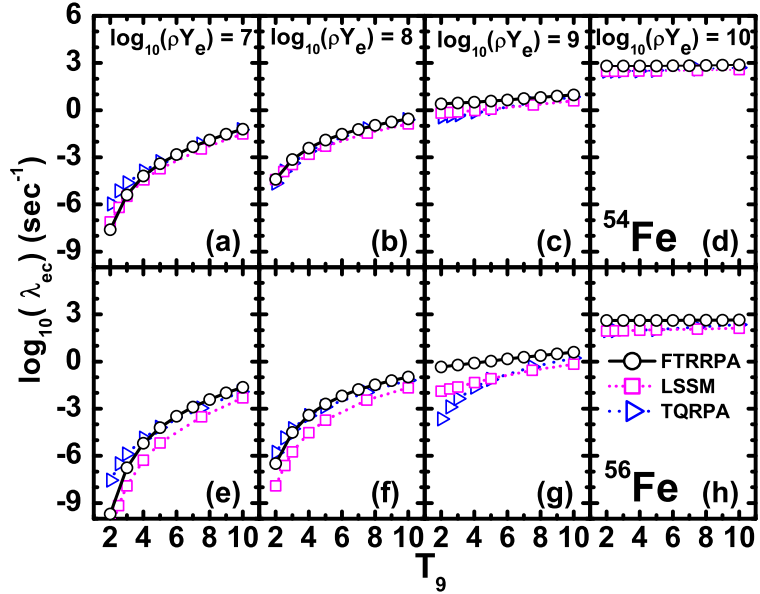}
 \caption{(Color online) Rates for electron capture on $^{54,56}$Fe
as functions of the temperature $T_9$ ($T_9=10^9$ K), at selected
densities $\rho Ye$ (g cm$^{-3}$). The results calculated using the
FTRRPA with the DD-ME2 effective interaction, are shown in
comparison with the rates obtained with LSSM calculations
\cite{Langanke2001data} and the TQRPA model \cite{Dzhioev2010}.}
\label{fig6}
\end{figure}

Figure~\ref{fig6} shows the calculated rates for electron capture on
$^{54,56}$Fe as functions of the temperature $T_9$ ($T_9=10^9$ K),
selected densities $\rho Y_e$ (g cm$^{-3}$). For comparison with the
FTRRPA results, the rates obtained with LSSM
calculations~\cite{Langanke2001data} and the TQRPA
model~\cite{Dzhioev2010} are also included in the figure. Here only
the 1$^+$ transitions are included in the calculation of cross
section. Although the three models compared here are based on rather
different assumptions, the resulting capture rates nevertheless show
similar trends. In general the electron-capture rates increase with
temperature and electron density. For high electron densities the
rates increase slower, and at density $\rho Y_e=10^{10}$ g/cm$^3$
the temperature dependence almost vanishes. At high densities
characterized by large values of the electron chemical potential,
high-energy electrons excite most or even all the GT$^+$ transitions
independent of temperature. Under such condidtions the increase in
temperature will not have a pronounced effect on the capture rates.
By increasing the number of neutrons from $^{54}$Fe to $^{56}$Fe,
one notices that the capture rates are slightly reduced in
$^{56}$Fe, reflecting the behavior of the cross sections.

The FTRRPA results generally reproduce the temperature dependence of
the rates predicted by the LSSM, but on the average the values
calculated with the FTRRPA are somewhat larger, especially for
$^{56}$Fe. For $^{54}$Fe and at lower densities $\rho Y_e = 10^7$ or
$10^8$ g/cm$^3$, the FTRRPA results essentially coincide with the
shell model calculation. At higher density, e.g. $\rho Y_e = 10^9$
g/cm$^3$, and with the electron chemical potential $\approx 5$ MeV
close to the threshold energy, the FTRRPA yields higher rates at
lower temperature. One can understand this difference from the
fragmentation of the shell model GT$^+$ strength over the energy
range $0\sim10$ MeV~\cite{Caurier1999}. While electrons at lower
temperature excite all the GT$^+$ strength in FTRRPA (see
Fig.~\ref{fig1}), only a fraction of the shell-model strength is
excited. Because part of the shell-model GT$^+$ strength is located
at higher energies than in the FTRRPA calculation, the resulting
LSSM rates are smaller. At even higher densities, e.g. at $\rho Y_e
= 10^{10}$ g/cm$^3$ with the chemical potential $\approx 11$ MeV,
already at lower temperatures the high-energy electrons excite all
the transition shell-model strength, and the resulting rates are
essentially the same as those calculated with the FTRRPA. For
electron capture on $^{56}$Fe, at lower densities $\rho Y_e = 10^7$
and $10^8$ g/cm$^3$ the FTRRPA results are in better agreement with
the TQRPA calculation, whereas the LSSM predicts lower rates. At
higher densities the trend predicted by the FTRRPA is closer to the
LSSM, but the calculated values are still above the shell model
results. In general, the differences between the FTRRPA and the
shell-model rates are larger in $^{56}$Fe than $^{54}$Fe. The
principal reason lies in the difference between the GT$^+$ centroid
energies calculated in the two models (cf. see Fig.~\ref{fig1}). As
in the case of $^{54}$Fe, the largest difference between the FTRRPA
and LSSM is at $\rho Y_e = 10^9$ g/cm$^3$, because the electron
chemical potential at this density is close to the threshold energy,
hence the capture rates are sensitive to the detailed GT$^+$
distribution.
\begin{figure}
\centerline{
\includegraphics[scale=0.45,angle=0]{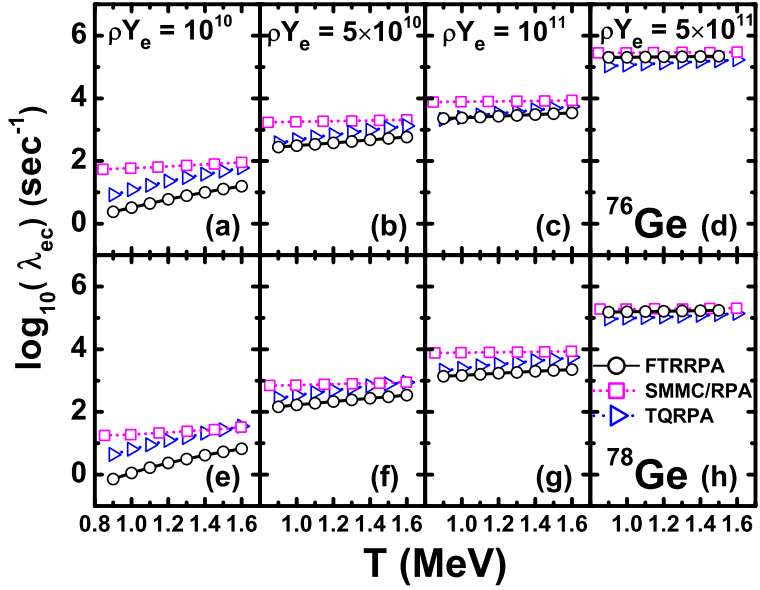}
} \caption{(Color online) Rates for electron capture on $^{76,78}$Ge
as functions of the temperature, at selected densities $\rho Ye$ (g
cm$^{-3}$). The results calculated using the FTRRPA with the DD-ME2
effective interaction, are shown in comparison with the rates
obtained with the hybrid model (SMMC/RPA) and the TQRPA
model~\cite{Dzhioev2010}.} \label{fig7}
\end{figure}

Fig.~\ref{fig7} compares the rates for electron capture on
$^{76,78}$Ge, calculated using the FTRRPA with the DD-ME2 effective
interaction, to the values obtained with the hybrid model (SMMC/RPA)
and the TQRPA model~\cite{Dzhioev2010}. In order to allow a direct
comparison with the hybrid model, the same quenching of the
axial-vector coupling constant with respect to its free-nucleon
value is employed, i.e. $g_A^*$=0.7$g_A$. Because for $^{76,78}$Ge
the contribution of forbidden transition is not negligible, the
calculations of rates Eq.~(\ref{ec_rate}) includes the multipole
transitions $J^\pi=0^{\pm},1^\pm, 2^\pm$. Similar to the case of Fe
nuclei, the calculated capture rates increase with temperature and
density, and are reduced by adding neutrons from $^{76}$Ge to
$^{78}$Ge. For $^{76,78}$Ge the rates predicted by the FTRRPA
display a temperature and density dependence very similar to that of
the TQRPA model, whereas the hybrid model predicts a very weak
temperature dependence at all densities considered in
Fig.~\ref{fig7}. In general, both the FTRRPA and the TQRPA predict
smaller values of capture rates compared to the hybrid model. The
reason is that the probability of unblocking transition channels is
larger in the hybrid model because it includes many-body
correlations beyond the RPA level. At the density $\rho Ye =
10^{10}$ g/cm$^3$ the FTRRPA capture rates exhibit a relatively
strong temperature dependence. The electron chemical potential is
$\approx 11$ MeV, and the cross sections are dominated by GT$^+$
transitions. By increasing temperature the GT$^+$ transitions are
unblocked, resulting in a large enhancement of the cross sections as
shown in Fig.~\ref{fig5}. A similar trend is also predicted by the
TQRPA calculation~\cite{Dzhioev2010}, whereas the temperature
dependence of the capture rates is much weaker in the hybrid model.
With a further increase in density to $\rho Ye = 10^{11}$ g/cm$^3$,
the chemical potential reaches $\approx 23$ MeV. At these energies
forbidden transitions dominate the calculated cross sections, the
FTRRPA yields cross sections similar to the TQRPA, and the same for
the capture rates. At even higher densities the temperature
dependence of the FTRRPA and TQRPA results becomes weaker, because
the cross sections are less sensitive to temperature. At the density
$\rho Ye = 5 \times 10^{11}$ g/cm$^3$ the capture rates predicted by
the FTRRPA are larger than the TQRPA results, and reach values
similar to those of the hybrid model.

\section{Conclusion}

In this work we have introduced a self-consistent theoretical
framework for modelling the process of electron capture in the
initial phase of supernova core collapse, based on relativistic
energy density functionals. The finite-temperature RMF model is
employed to determine the single particle energies, wave functions
and thermal occupation probabilities for the initial nuclear states.
The relevant charge-exchange transitions $J^{\pi} = 0^{\pm}$,
$1^{\pm}$, $2^{\pm}$ are described by the finite-temperature
relativistic random-phase approximation (FTRRPA). The FTRMF+FTRRPA
framework is self-consistent in the sense that the same relativistic
energy density functional is employed both in the finite-temperature
RMF model and in the RRPA matrix equations.

In the calculation of the electron capture cross sections, the
GT$^+$ transitions provide the major contribution in the case of
$^{54,56}$Fe, whereas for more neutron-rich nuclei such as
$^{76,78}$Ge forbidden transitions play a more prominent role
already starting at incident electron energy above $\approx$10 MeV.
The principal effect of increasing temperature is the lowering of
the electron-capture threshold energy. For $^{76,78}$Ge the cross
sections in the low-energy region are sensitive to temperature
because of the dominant role of GT$^+$ transition channel, but these
correlation becomes weaker at higher energies dominated by major
contributions from forbidden transitions.

Electron capture rates for different stellar environments, densities
and temperatures, characteristic for core collapse supernovae have been
calculated and compared with previous results of
shell-model, hybrid shell-model plus RPA, and thermal QRPA (TQRPA)
calculations. For $^{54,56}$Fe, the FTRRPA results generally reproduce the
temperature dependence of the rates predicted by shell-model calculations, but
on the average the values calculated with the FTRRPA
are somewhat larger, especially for $^{56}$Fe.
For $^{76,78}$Ge the FTRRPA capture rates display a trend very similar
to that of the TQRPA calculation, especially for the temperature dependence,
whereas this dependence of the capture rates is much weaker in the hybrid model.

The results obtained in the present study demonstrate that the
framework of finite-temperature RMF and FTRRPA provides a universal
theoretical tool for the analysis of stellar weak-interaction
processes in a fully consistent microscopic approach. This is
especially important for regions of neutron-rich nuclei where the
shell-model diagonalization approach is not feasible. A microscopic
approach has a big advantage in comparison to empirical models that
explicitly necessitate data as input for calculations, as in many
mass regions data will not be available. Of course, the present
framework is limited to the level of RPA and does not include
important many-body correlations that are taken into account in a
shell-model approach. However, as discussed previously, at higher
densities and temperatures in the stellar environment, the detailed
fragmentation of transition spectra does not play such a significant
role, and the FTRRPA represents a very good approximate framework
that can be used in systematic calculations of electron-capture
rates. Further improvements of the current version of the model are
under development. For open-shell nuclei at very low temperatures,
pairing correlations need to be taken into account. To obtain the
empirical fragmentation of the transition spectra, the inclusion of
higher-order correlations beyond the RPA level, that is, the
coupling to $2p - 2h$ states will be  necessary.

{\center{\bf ACKNOWLEDGMENTS}}

Y.F. Niu would like to acknowledge discussions with Z. M. Niu and H.
Z. Liang. This work is supported by the State 973 Program
2007CB815000, the NSF of China under Grants No. 10975007 and No.
10975008, the Unity through Knowledge Fund (UKF Grant No. 17/08),
and by MZOS - project 1191005-1010, and the Chinese-Croatian project
``Nuclear structure and astrophysical applications".

%

\clearpage


\end{document}